\documentstyle[aps]{revtex}
\begin{document}
\draft
\author{V.F. Dmitriev\footnote{e-mail address: dmitriev@inp.nsk.su},
I.B. Khriplovich\footnote{e-mail address: khriplovich@inp.nsk.su}
and V.B. Telitsin\footnote{e-mail address: telitsin@inp.nsk.su}}
\address{Budker Institute of Nuclear Physics,\\
 Novosibirsk-90, 630090, Russia}
\date{\today}
\title{Is Large Weak Mixing in Heavy Nuclei\\
Consistent with Atomic Experiments?}
\maketitle

\begin{abstract}
The hypothesis of a large weak matrix element between single-particle
states in heavy nuclei ($\sim 100$ eV) contradicts
the results of atomic PNC experiments.
\end{abstract}

The scattering cross-sections of longitudinally polarized epithermal
(1 - 1000 eV) neutrons from heavy nuclei at $p_{1/2}$ resonances have
large longitudinal asymmetry. This parity nonconserving (PNC)
correlation is the fractional difference of the resonance
cross-sections for positive and negative neutron helicities. For a
long time the most natural explanation of the effect was based on the
statistical model of the compound nuclei. In fact, not only the
explanation, but the very prediction of the huge magnitude of this
asymmetry (together with the nuclei most suitable for the
experiments) was made theoretically \cite{sf} on the basis of this
model.

An obvious prediction of the statistical model is that after
averaging over resonances, the asymmetry should vanish. However, few
years ago it was discovered \cite{fb,fb1} that all seven asymmetries
for $^{232}$Th have the same, positive sign. This tendency was observed
also in other nuclei.

All the attempts \cite{bg,kj,ab,lw} to explain a common sign require
the magnitude of the weak interaction matrix element, mixing
opposite-parity nuclear levels, to be extremely large, $\sim 100$ eV.
The same assumption seems to be necessary to explain unexpectedly
large P-odd correlations observed in M\"{o}ssbauer transitions in
$^{119}$Sn and $^{57}$Fe \cite{ts,ts1}.

In a recent paper \cite{sb} it was pointed out that such a large
magnitude of the weak mixing can be checked in an independent
experiment. The proposal is to measure PNC asymmetry in the M4
$\gamma$-transition between the (predominantly) single-particle
states $1i\; 13/2^+$ and $2f\; 5/2^-$ in $^{207}$Pb. The experiment
sensitivity to the weak matrix element value is expected to reach
$5 - 13$ eV.

In the present Comment we wish to note that close upper
limit on the weak mixing in $^{207}$Pb can be extracted now from the
measurements of the PNC optical activity of atomic lead vapour
\cite{mvm}. The experiment was performed at the atomic M1 transition
from the ground state $6p^2\;^3P_0$ to the excited one $6p^2\;^3P_1$.
The nuclear spin of $^{207}$Pb being $i=1/2$, the total atomic angular
momentum of the ground level is $F=1/2$, and the upper level is split
into two: $F^{\prime}=1/2,\,3/2$. The following upper limit was
established at the 95\% confidence level for the relative magnitude of
the nuclear-spin-dependent (NSD) part of the optical activity:
\begin{equation}\label{ra}
\frac{P_{NSD}}{P} < 0.02
\end{equation}
Here
$$P_{NSD}
= P(F=1/2\rightarrow F^{\prime}=1/2) - P(F=1/2\rightarrow F^{\prime}=3/2)$$
and $P$ is the main, nuclear-spin-independent, part of the PNC
optical activity.

In heavy atoms the NSD P-odd effects were shown to be induced mainly
by contact electromagnetic interaction of electrons with the anapole
moment of a nucleus which is its P-odd electromagnetic characteristic
induced by PNC nuclear forces \cite{fk,fks}.

The electromagnetic PNC interaction of electrons with nuclear AM is
of a contact type. It is conveniently characterized in the units of
the Fermi weak interaction constant $G=1.027\times 10^{-5} m^{-2}$
($m$ is the proton mass) by a dimensionless constant $\kappa$.

To calculate $\kappa$ let us present the effective P-odd potential
for an external nucleon in a contact form in the spirit of the
Landau-Migdal approach:
\begin{equation}\label{we}
W=\frac{G}{\sqrt{2}}\;\frac{g}{2m}\;\vec{\sigma}[\vec{p}\rho(r)
+\rho(r)\vec{p}\;].
\end{equation}
Here $\vec{\sigma}$ and $\vec{p}$ are respectively spin and momentum
operators of the valence nucleon, $\rho(r)$ is the density of
nucleons in the core normalized by the condition $\int
d\vec{r}\rho(r)=A$ (the atomic number is assumed to be large, $A\gg
1$). A dimensionless constant $g$ characterizes the strength of the
P-odd nuclear interaction.  It is an effective one and includes
already the exchange terms for identical nucleons. This constant
includes also additional suppression factors reflecting long-range
and exchange nature of the P-odd one-meson exchange, as well as the
short-range nucleon-nucleon repulsion.

Under some simplifying assumptions the anapole constant $\kappa$ can
be estimated for a heavy nucleus even analytically with the following
result \cite{fks}:
\begin{equation}
\kappa=\frac{9}{10}\,g\,\frac{\alpha \mu}{m r_0}\, A^{2/3}.
\end{equation}
Here $\mu$ is the outer nucleon magnetic moment, $r_0=1.2$fm. The
enhancement $\sim A^{2/3}$ compensates to a large extent the small
fine structure constant $\alpha=1/137$. That is why the nuclear AM is
perhaps the main source of the nuclear-spin-dependent PNC effects in
heavy atoms \cite{fk,fks}. This formula predicts for lead
\begin{equation}
\kappa(^{207}\makebox{Pb}) = - 0.08\;g_n.
\end{equation}
More serious numerical calculations using a realistic description of
the core density and a Woods-Saxon potential including the spin-orbit
interaction give \cite{fks,dkt}
\begin{equation}
\kappa(^{207}\makebox{Pb}) = - 0.105\; g_n.
\end{equation}

On the other hand, atomic calculations predict the magnitude of the
NSD optical activity in lead at given $\kappa$ with the accuracy
about 20\% \cite{nsfk,k}. At the experimental value of $P$ obtained
in Ref. \cite{mvm} this prediction for the ratio (\ref{ra})
constitutes
$\;\; 0.023\, \kappa(^{207}\makebox{Pb}).\;\;$
Combining the experimental result (\ref{ra}) with this theoretical one,
we get the following upper limit for the anapole constant:
\begin{equation}
\kappa(^{207}\makebox{Pb}) < 1,
\end{equation}
and for the effective neutron PNC constant:
\begin{equation}\label{co}
g_n < 10.
\end{equation}

Close upper limits on the effective constant $g_p$ for an outer
proton can be extracted from the optical experiments with atomic
cesium \cite{wi} and thallium \cite{la}. Less strict bound on $g_p$
follows from the experiment \cite{mzw} with bismuth.

A simple-minded estimate for the weak mixing matrix element, based on
formula (\ref{we}), leads to its following value:
\begin{equation}
< W >\, \simeq \, 2\,g\;eV.
\end{equation}
More sophisticated calculations based on a Woods-Saxon potential with
the spin-orbit interaction gives for the concrete matrix element of
interest for the proposed experiment with $^{207}$Pb
\begin{equation}\label{ca}
\langle 3d\; 5/2^+|W| 2f\; 5/2^-\rangle = 1.4\,g_n\;eV
\end{equation}
in a reasonable agreement with the results of other single-particle
nuclear calculations cited in Ref. \cite{sb}. Combining (\ref{co})
and (\ref{ca}), we get the following upper limit on this matrix
element
\begin{equation}
\langle 3d\; 5/2^+|W| 2f\; 5/2^-\rangle < 14 \; eV
\end{equation}
which is close to the expected accuracy of the experiment discussed
in Ref. \cite{sb}. Nevertheless, this experiment would be obviously
both interesting and informative, so much the more that it would be
the first occasion when PNC effects in the same nucleus were measured both
in atomic and nuclear experiments.

As to the hypothesis itself, according to which the magnitude of the weak
mixing matrix element is as high as 100 eV, such a large its
value does not agree with the results of atomic PNC experiments.

\bigskip
\bigskip
This investigation was financially supported by the Russian
Foundation for Basic Research, grant No. 94-02-03942-a.



\end{document}